\documentclass{comjnl}

\usepackage[cmex10]{amsmath}
\usepackage{amssymb}
\usepackage{graphicx}
\usepackage{figsize}
\usepackage{float}
\usepackage{verbatim}
\usepackage{array}
\usepackage{url}

\begin{document}

\title[Generating and evaluating application-specific hardware extensions]{Generating and evaluating application-specific hardware extensions}
\author{Nikolaos~Kavvadias}
\affiliation{Department of Computer Science and Technology,
University of Peloponnese, 
Terma Karaiskaki,
Tripoli 22100, Greece} \email{nkavv@uop.gr}
\shortauthors{N. Kavvadias}

\received{00 December 2009}
\revised{00 Month 2010}


\begin{abstract}
Modern platform-based design involves the application-specific extension of embedded processors to fit customer requirements. To accomplish this task, the possibilities offered by recent custom/extensible processors for tuning their instruction set and microarchitecture to the applications of interest have to be exploited. A significant factor often determining the success of this process is the utomation available in application analysis and custom instruction generation. 

In this paper we present YARDstick, a design automation tool for custom processor development flows that focuses on generating and evaluating application-specific hardware extensions. YARDstick is a building block for ASIP development, integrating application analysis, custom instruction generation and selection with user-defined compiler intermediate representations. In a YARDstick-enabled environment, practical issues in traditional ASIP design are confronted efficiently; the exploration infrastructure is liberated from compiler and simulator idiosyncrasies, since the ASIP designer is empowered with the freedom of specifying the target architectures of choice and adding new implementations of analyses and custom instruction generation/selection methods. To illustrate the capabilities of the YARDstick approach, we present interesting exploration scenarios: quantifying the effect of machine-dependent compiler optimizations and the selection of the target architecture in terms of operation set and memory model on custom instruction generation/selection under different input/output constraints.
\end{abstract}

\maketitle

\section{Introduction}
\label{Sec:Intro}
{ASIPs} (\emph{Application Specific Instruction-set Processors}) play a central role in contemporary embedded systems-on-a-chip (SoCs) replacing hardwired solutions which offer no programmability for enabling reuse or encompassing late specification changes. ASIPs are tuned for cost-effective execution of targeted application sets. An ASIP design flow involves profiling, architecture exploration, generation/selection of instruction-set extensions (ISEs) and synthesis of the corresponding hardware while enabling the user taking certain decisions. 

Custom processors either adhere to the configurable/extensible processor paradigm \cite{ARC,Gonzalez00} or can be ASIPs completely designed from scratch. Configurability lies in tuning architectural parameters (e.g. cache sizes) and enabling/disabling features \cite{Yiannacouras06} while extensibility of a processor comes in modifying the instruction set architecture by adding forms of custom functionality. Designing a custom processor from scratch is a more aggressive approach requiring a significant investment of effort in developing all the necessary software development tools (compiler, binary utilities, debugger/simulator) and possibly a real-time OS, while in the configurable processor case, the RTOS is usually targeted to the base ISA and the software toolchain can be incrementally updated. There exist two basic themes for architecture extension: tight integration of custom functional units and storage \cite{NiosII} or loose coupling of hardware accelerators 
to the processor through a bus interface \cite{MicroBlaze}. Recent works \cite{Sirowy07} 
advocate in favor of both approaches, proving that both techniques can be considered 
simultaneously by formalizing the problem as a form of two-level partitioning. 

It is often in ASIP/custom processor design that certain practical issues arising from seemingly invariant elements of the design flow are not addressed:
\begin{enumerate}
\item[a)] {Assumptions of the intermediate representation (IR) to which the application code is mapped, directly affect solution quality as in the case of ISE synthesis.} 
\item[b)] {The exploration infrastructure tied up to the conventions of software development tools.} 
\item[c)] {Adaptability to different compilers/simulators.}
\item[d)] {Support for low-level entry for application migration within a processor family and reverse engineering.}
\end{enumerate}

In this paper, all these issues are successfully addressed by integrating custom instruction (CI) generation and selection techniques with a flexible IR infrastructure that can reflect certain designer decisions that is cumbersome to apply otherwise. Our approach is substantiated in the form of the YARDstick prototype tool \cite{Kavvadias07}. For example using an IR with intrinsic support for bit-level operations may yield significantly different ISEs to the case of an unaugmented IR. Also, in YARDstick it is possible to directly measure the effect of certain machine-dependent compiler transformations, such as register allocation, to the quality and impact of the generated ISEs, an issue recognized but never quantified in other works \cite{ClarkN03,Castro04}. Further, YARDstick provides profiling facilities for determining static and dynamic application metrics such as data types, memory hierarchy statistics, and execution frequencies. Application entry can be either high-level (e.g. ANSI C) or low-level (assembly code for a target architecture or virtual machine). A number of recent custom functionality identification and selection techniques have been implemented while hardware estimators (speedup, area) and bindings to third-party tools for hardware synthesis from CDFGs are provided. 

It is important to note that the interpretation of custom functionalities depends on the context; they can represent instruction-set extensions (ISEs) to a baseline ISA requiring to be accounted in the control path of the processor (decoding logic, extending the interrupt services), custom instructions of an ASIP enabled by a programmable controller or hardwired functions meant to be used as non-programmable hardware accelerators, loosely connected to the processor (i.e. accessible through the local bus).

\section{Related work}
\label{Sec:RelatedWork}
Last years, a number of research efforts have regarded the automated application-specific 
extension of embedded processors \cite{Alippi99,Yu04a,ClarkN05,Goodwin03,Pozzi06,Biswas07,Pothineni07}. 
A few open instruction generation frameworks exist \cite{Pattlib}; an 
advantage of their work being delivering a format for storing, manipulating and 
exchanging instruction patterns. In order to use their pattern library (Pattlib), the 
potential user should adapt his compiler for generating and manipulating patterns in 
the cumbersome GCC RTL (Register Transfer Language) \cite{GCC} intermediate representation. Some issues with the Pattlib approach regard the significant efforts for adapting the GCC compiler to emit information in ``pattlib'' format, and that the IR for their selected backend (SPARC V8) is not architecture-neutral and cannot be easily altered. 
 
Application-specific instructions have been generated for the Xtensa configurable 
processor \cite{Goodwin03} that may comprise of VLIW (Very Long Instruction Word), SIMD 
(Single-Instruction Multiple-Data) or fused (chained) RTL operations. However, as induced by the architecture template of Xtensa, control-transfer instructions ({\it cti}) are not considered to be included in the resulting complex instructions. A sophisticated framework for the design of tightly-coupled custom coprocessing datapaths and their integration to existing processors has been presented in \cite{ClarkN05}. While providing a complete solution to programmable acceleration, their work still has some drawbacks: the possibility 
of direct communication to fast local data memory is excluded and for this reason, 
beneficial addressing modes cannot be identified. In \cite{Atasu03,Pozzi06} a multi-output 
instruction generation algorithm is presented which selects maximal-speedup convex 
subgraphs for each basic block data-dependence graph (DDG), with worst case exponential 
complexity, while \cite{Yu04a} added path profiling to extend beyond basic block scope.
An important conclusion was that useful instruction identification scope does not extend 
further than 2 or 3 consecutive basic blocks. Still, memory operations are not regarded 
in the formation of custom instructions, while pattern identification can only take 
place post register allocation.

\section{YARDstick}
\label{Sec:YARDstick}
The main role of YARDstick is to facilitate design space exploration (DSE) in heterogeneous flows for ASIP design where the development tools (compiler, binary utilities, simulator/ debugger) in many cases, lack DSE capabilities and/or have been designed with different interfaces in mind. Thus, it is often that significant development effort is required in adding features as afterthoughts and dealing with interoperability issues, especially at the {\it compiler} and {\it simulator} boundaries.

\subsection{The YARDstick kernel}
\label{Sec:YARDstickKernel}
\begin{figure}[tb]
  \centering
  \includegraphics[width=8.0cm]{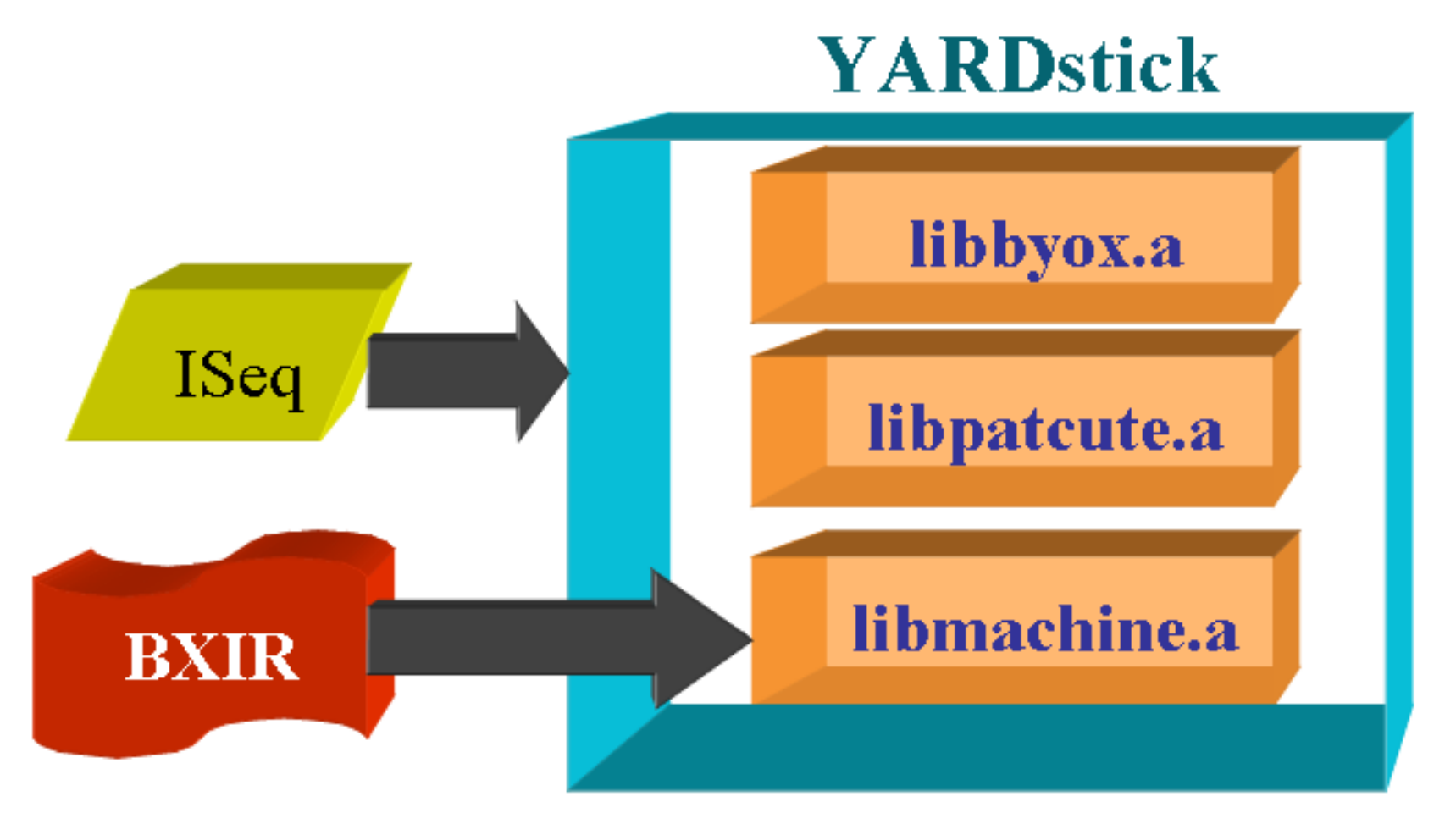}
  \caption{The YARDstick infrastructure.}
  \label{Fig:1}
  \vspace{-0.25cm}
\end{figure}

The current YARDstick infrastructure, as illustrated in Fig.~\ref{Fig:1}, comprises of three kernel components ({\it libByoX, libPatCUTE, libmachine}), the target architecture specification tools (the BXIR frontend) and a set of backends for exporting control-flow graphs, basic blocks and custom instructions for visualization, simulation and RTL synthesis purposes. {\it libByoX} and {\it libPatCUTE} are target-independent, and only {\it libmachine} has to be retargeted for different IR specifications.

\subsubsection{{\it libByoX}}
\label{Sec:libbyox}
{\it libByoX} implements the core YARDstick API and provides frontends/manipulators for internal data structures. 
The ByoX (Bring Your Own Compiler and Simulator) library provides:
\begin{itemize}
\item {The ISeqinfo parser for ISeq (historical name for ``Instruction Sequence'') entries. ISeq is a flat CDFG (with/without SSA) format of application IR that is used for recording the data-dependence graphs for the application basic blocks.}
\item {The CFGinfo parser for control-flow graph (CFG) files that attribute the corresponding ISeq files with typed control-flow edges.}
\item {Simple file interface for the ISeq and CFG formats as well as for results of compiler analyses, e.g. control/data flow analyses evaluating register liveness and natural loops, that can be passed to ByoX as defined by their corresponding BNFs.}
\item {An IR manipulation API for writing external analyses and optimizations.}
\item {Parameterization for a template machine context without inherent restrictions to its ISA.}
\end{itemize}

In ISeq, the following application information is recorded:
\begin{itemize}
\item	{The global symbol table.}
\item {The procedure list, consisting of data dependence entries, the local symbol table and a statement list per procedure. It is possible to generate different facets of the local symbol table, e.g. single reference per direction (input or output) for each variable versus allowing multiple definition points for the same variable.}
\end{itemize}

\subsubsection{{\it libPatCUTE}}
\label{Sec:libpatcute}
Further, a number of custom instruction generation/selection methods have been implemented as part of the PatCUTE (Pattern-based Custom UniT Exploration) library. CI generation involves the identification of MIMO (Multiple-Input Multiple-Output) or MISO (Multiple-Input Single-Output) ISeq patterns under user-defined constraints. The CI generation methods available in {\it libPatCUTE} are:
\begin{itemize}
\item {MAXMISO \cite{Alippi99} for identifying maximal subgraphs with a single-output node using a linear complexity algorithm.}
\item {MISO exploration under constraints for the maximum number of input/output operands, and for two types of operation node-related constraints \cite{Kavvadias05}.} 
\item {MIMO CI generation. In our case, we do not search for maximal MIMO patterns \cite{Pothineni07}, however, we employ a fast heuristic by assuming similarly to \cite{Pothineni07} that the performance gain provided by a pattern $P$ is higher than any pattern that is a subgraph of $P$. The user could disable the heuristic and apply an exponential complexity algorithm as well.}
\end{itemize}

When CI generation is invoked, a CI list is constructed from the resulting ISeq patterns, which can be filtered via graph or graph-subgraph isomorphism tests \cite{VFLib2} during the process of removing redundant cases. A subset of the library can be selected by using either a configurable greedy selector (supporting cycle-gain and cycle-gain per area priority metrics) or a 0-1 knapsack-based one. An important YARDstick characteristic is that CIs can be expressed in ISeq in the same way to either application CFGs or subregions thereof, thus existing data structures and analyses can be reused for further manipulation of the generated CIs. For example, pattern libraries can be imported to YARDstick.

\subsubsection{{\it libmachine}}
\label{Sec:libmachine}
The {\it libmachine} library is the only core YARDstick component that needs retargeting for a user-defined target architecture. Target architectures are specified in the BXIR (ByoX IR) 
format which supports semantics for defining global-scope (data types, operation grouping) and operation-level information (operands, interpretation semantics for each IR operator, area/latency cost for corresponding hardware implementations and cycle timings). 

\subsubsection{Backend engines}
\label{Sec:backends}
Application CFGs, (basic blocks) BBs and patterns can be processed by a number of backends for exporting to:
\begin{itemize}
\item {ANSI C subset code for incorporation to user tools (simulators, validators etc).}
\item {GDL (VCG) \cite{VCG} and dot (Graphviz) \cite{Graphviz} files for visualization.}
\item {An extended CDFG \cite{CDFGtool} format for scheduling and translation to synthesizable VHDL (applicable to BBs and CI patterns).}
\item {GGX XML \cite{AGG} files for algebraic graph transformation.}
\end{itemize}

\subsection{Structure of a YARDstick environment}
\label{Sec:YARDstickStructure}
\begin{figure}[tb]
  \centering
  \includegraphics[width=8.0cm]{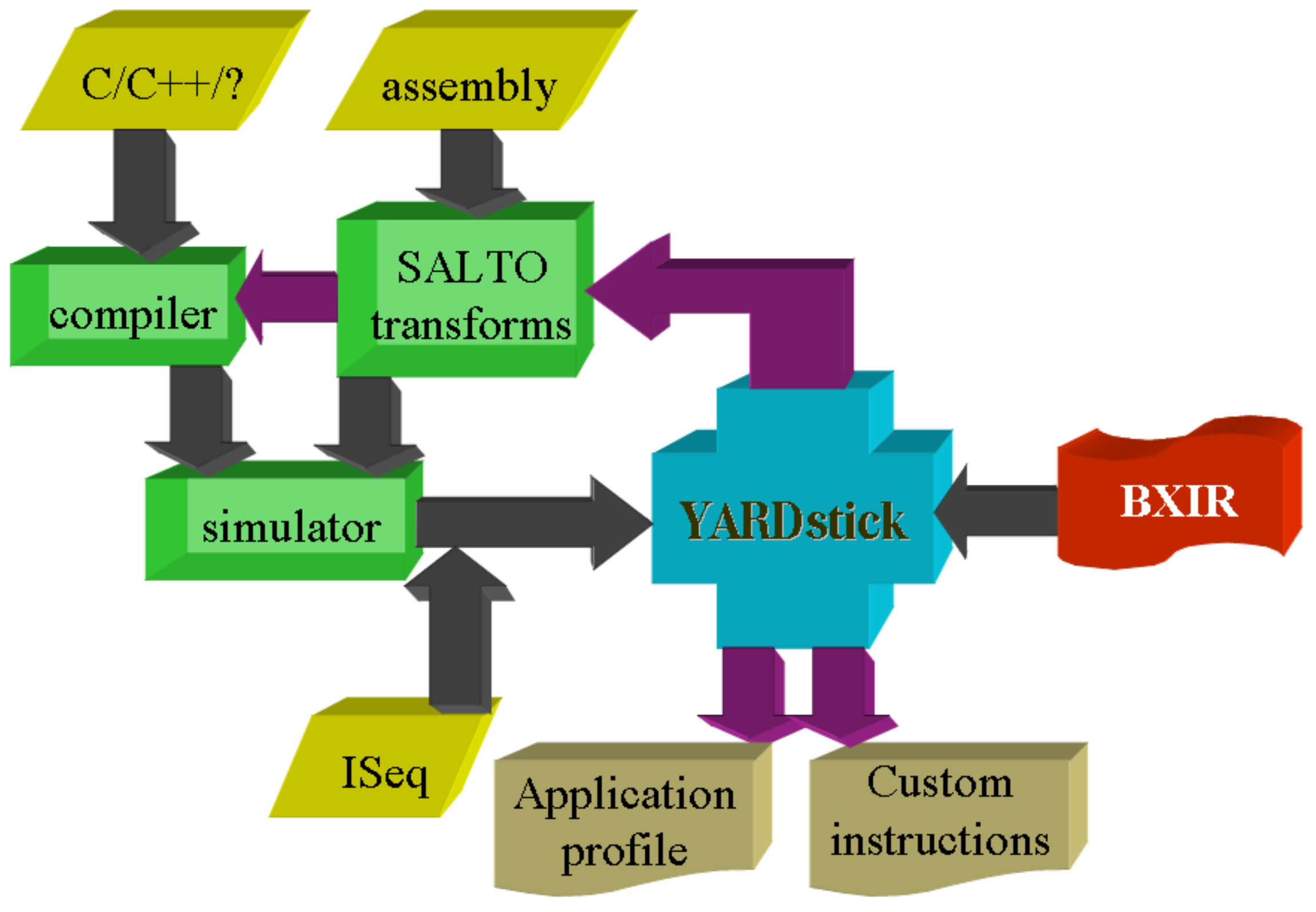}
  \caption{A high-level look to a YARDstick-based framework.}
  \label{Fig:2}
  \vspace{-0.25cm}
\end{figure}

The YARDstick kernel can be utilized as an infrastructure for application analysis and exploration of custom functionality extensions. Fig.~\ref{Fig:2} shows our YARDstick framework which reuses third-party compilation and simulation tools. The compiler frontend ({\it gcc} is such an example) accepts input in C/C++ or other high-level languages of interest. The application program is compiled to a low-level representation that can be represented by a form of ``assembly'' code after frontend processing, conversion to its internal IR, application of machine-independent optimizations and a set of compiler backend processes with only code selection being obligatory. The assembly-level code can then be macro-expanded, instrumented for profiling and converted to ISeq by an appropriate SALTO pass \cite{SALTO}. This flow assumes that a working SALTO backend library has been ported for the target architecture. Assembly code can be assembled and linked by the target machine binary utilities ({\it binutils} or equivalent tools) and the resulting ELF executables can be evaluated on an instruction- or cycle-accurate simulator. Alternatively, ISeq files can be generated as compiler IR dumps directly from the compiler for the target machine. This is the case for a modified version of Machine-SUIF \cite{MachSUIF} for which the basic block profile is automatically obtained by converting the IR to a C subset and executing the low-level C code on a native machine.

At the simulation boundary, YARDstick expects information on the dynamic profile of the application (basic block execution frequencies, program trace, cache memory access statistics) on a target machine. From within YARDstick, static and dynamic application metrics can be evaluated and visualized. An application analyzer ({\it iseqtool}) and CI generator ({\it igensel}) linked to {\it libByoX} and {\it libPatCUTE} are used to obtain the application profile and custom instructions, respectively. 

\subsection{Usage of the YARDstick API}
\label{Sec:APIUsage}
The YARDstick API provides methods for manipulation of ISeq entities and extraction of useful information to internal data structures such as local operand lists, operation-level 
analysis (e.g. finding zero-predecessor/-successor instruction nodes) and application of backend processing. Fig.~\ref{Fig:3} shows an example of API usage for updating the necessary data structures for basic-block-based CI generation. 

In more detail, a basic block ISeq cluster is denoted by `bb'. First, `init\_opnd\_library' initializes an empty operand list container, named {\it Lopnd} which is updated by calls to the `find\_{\it type}\_opnds' functions, where {\it type} denotes operand type and can be one of \{input,output,cnst\}. When the unique register operands and constants option is enabled, operands are treated as in SSA form and have a single representation per input and output sublist by applying `collapse\_to\_unique\_opnds'. After clearing temporary storage for the best cut to be identified in the specific iteration of the CI generation algorithm, either a MIMO or a MISO-based method can be selected for performing the actual process.

\begin{figure}[tb]
\centering
{
{\footnotesize
\begin{verbatim}
void evaluate_bb_ci(ISeq bb)
{
  ... 
  // Setup operand list
  UIOCList Lopnd = init_opnd_library();

  // Find unique i/o registers and constants
  find_input_opnds(bb, Lopnd, input_opnds);
  find_output_opnds(bb, Lopnd, output_opnds, 
  instr_has_successor);
  find_cnst_opnds(bb, Lopnd, cnst_opnds);

  if (unique i/o instances for operands/constants)
    collapse_to_unique_opnds();
  
  clear_best_cut();
  
  // CI generation for the BB
  if (MIMO method)
    MIMO_identification(bb);
  else if (MaxMISO or constrained MISO method)
    MAXMISO_identification(bb);
}
\end{verbatim}
}
}
\vspace{-0.125cm}
\caption{Updating internal data structures for BB-level CI generation.}
\label{Fig:3}
\end{figure}

\section{Case studies of design space exploration with YARDstick}
\label{Sec:DSE}
For proof-of-concept, we have evaluated YARDstick under various scenarios that reflect realistic problems in evaluating and exploring the design space when developing new ASIPs or enhancing customizable architectures. For the case studies we have used three different target architectures: 
\begin{enumerate}
\item [1)] {The SUIFvm IR \cite{MachSUIF} augmented by a set of incremental extensions to it, called {\it SUIFvmenh}.}
\item [2)] {The {\it SUIFrmenh} architecture (SUIF `real machine' enhanced) supported by an in-house backend written for Machine-SUIF, that introduces a finite register set of configurable size to {\it SUIFvmenh}. {\it SUIFrmenh} also resolves type casting (conversion) operations mapping them from the {\it cvt} SUIFvm instructions to the proper instructions accepted by the {\it SUIFrmenh} backend: zero- and sign-extend, truncation and {\it mov} explicitly denoting the source and destination data types.}
\item [3)] {The DLX integer subset ({\it iDLX}) for which the formatted assembly dumps are viewed as a kind of human-readable machine-level IR.}
\end{enumerate}

The target IR architectures are summarized in Table~\ref{Tab:1}. In the experiments of the following subsections, all control transfer instructions ({\it beqz, bnez, j, jr, jal, jalr}) were forbidden from CI pattern formation for the {\it iDLX} IR, while for the SUIFvm-based IRs, branch operations were permitted. The two different forbidden instruction constraint sets were chosen in order to highlight distinct potential requirements and were not meant to be directly contrasted. The DLX-based IR would be a choice when the objective is to optimize pre-existing DLX legacy assembly (or binaries). Instead, {\it SUIFvmenh} implements a representative RISC-like IR not restricting the processor template, which is more suitable for developing ASIPs from scratch.  

\begin{table}
  \renewcommand{\arraystretch}{0.925}
  \vspace{-0.25cm}
  \caption{Different IR settings for CI generation.}
  \centering
  {\footnotesize
  \begin{tabular}{|l|l|}
    \hline
    \multicolumn{1}{|m{1.5cm}|}{\centering IR}
    &\multicolumn{1}{m{5.0cm}|}{\centering Operations}\\
    \hline
    SUIFvmenh & SUIFvm plus: type conversion (sxt, zxt),\\
    & partial predication (select),\\ 
    & bit manipulation (bitinsert, bitextract, concat) \\
    \hline
    SUIFrmenh & SUIFvmenh with finite register set \\
    & (12, 16, 32 or 64 registers); here 32 is used \\
    \hline
    iDLX & The DLX integer instruction set \\
    \hline
  \end{tabular}
  }
  \label{Tab:1}
\end{table}

For the experiments we used applications from a set of embedded benchmarks consisting of 5 cryptographic ({\it crc32, deraiden, enraiden, idea, sha}) and 5 media-oriented applications ({\it adpcm\_dec, adpcm\_enc, fir, fsme, mc}) which are shown in Table~\ref{Tab:2}. 

\begin{table}
  \centering
  \renewcommand{\arraystretch}{0.925}
  \caption{Summary of examined benchmarks.} 
  {\footnotesize
  \begin{tabular}{|l|l|}
    \hline
    \multicolumn{1}{|m{2.0cm}|}{\centering Benchmark}
    &\multicolumn{1}{m{5.0cm}|}{\centering Description}\\
    \hline
    {\it crc32} & Cyclic redundancy check \\ \hline
    {\it deraiden} \cite{Raiden} & Decoding raiden cipher \\ \hline
    {\it enraiden} \cite{Raiden} & Encoding raiden cipher \\ \hline 
    {\it idea} & IDEA cryptographic kernel \\ \hline
    {\it sha} & Secure Hash Algorithm producing an 160-bit \\ 
    & message digest for a given input \\ \hline
    {\it adpcmdec} & Adaptive Differential Pulse Code Modulation \\ 
    & (ADPCM) decoder \\ \hline
    {\it adpcmenc} & Adaptive Differential Pulse Code Modulation \\ 
    & (ADPCM) encoder \\ \hline
    {\it fir} & FIR filter \\ \hline
    {\it fsme} & Full-search motion estimation \\ \hline
    {\it mc} & Motion compensation \\ \hline
  \end{tabular}
  }
  \label{Tab:2}
\end{table}

\subsection{Effect of compilation specifics: Case study of media processing kernels}
\label{Sec:CaseStudy}
It has been argued recently \cite{Bonzini06} that traditional compiler transformations and the trivial solution of applying CI identification at the end of the optimization phase pipeline do not necessarily yield the best performance when targeting a custom processor. On the contrary, source code and IR-level transformations have to be especially tuned for exposing beneficial application-specific hardware extensions. 

In this subsection, the effect of the choice in compilation specifics is highlighted for popular case study applications: the ADPCM codec, an FIR filter, and typical implementations of motion estimation/compensation. We investigate specific effects that the compiler imposes when used for exploring the potential for custom instructions:
\begin {enumerate}
\item[a)] {The effect of register allocation on the quality of the generated CIs. For this purpose, we have targeted the {\it SUIFrmenh} backend. A 32-entry register file was assumed while the procedure calling convention for {\it SUIFrmenh} was the same to an in-house GCC-based DLX backend.}
\item[b)]	{The suitability of using a highly-optimizing (but aimed to general-purpose processors) compiler such as {\it gcc} targeted to DLX which is our case, against a well-known research compiler ({\it MachSUIF} targeted to SUIFvm) which has been extensively used for exploring the transformation space for new CIs.}
\end{enumerate}

For accounting only the true data dependencies amongst operations, it is necessary to remove all false dependencies. This can be achieved by a simple IR-level transformation pass (for example, such pass was implemented in the Machine-SUIF compiler for the {\it SUIFvmenh} target) which involves the use of the pseudocode of Fig.~\ref{Fig:4}. The algorithm in Fig.~\ref{Fig:4} can be used for an in-order instruction schedule, i.e. no backward data dependence edges exist within a basic block. For a given set of dependence edges $\bigcup\{(i \rightarrow j)_k\}$ between instructions $mi(i),mi(j)$ of instruction IDs $i,j$ respectively, the range $[i,j]$ is considered. The destination operands of machine instructions in range are iterated and compared to the operand ($opnd$) for which we want to remove all false dependencies with it as the data dependency. If $opnd$ is written at least once, a false dependency is identified (marked as $TRUE$) and the corresponding data dependence edge is annulled.

\begin{figure}[tb]
\centering
{
{\footnotesize
\begin{verbatim}
boolean is_false_dependency(BB* bb, InstrID mi_lpos, 
  mi_hpos, LOpnd opnd)
{
  boolean false_dependency_f = FALSE;
  ...
  // Iterate through the [mi_lpos..mi_hpos] range
  foreach machine instruction (mi) in range do     
    if the current mi is within the specified range
      get destination operand dstop of mi      
      if dstop is ((a base register or address symbol) 
      and writes memory)
        if dstop is equal to opnd
          // a false dependency has been found
          false_dependency_f |= TRUE;
        fi
      fi
    fi
  od
  
  return false_dependency_f;
}
\end{verbatim}
}
}
\vspace{-0.125cm}
\caption{Removing false data dependence edges from basic blocks.}
\label{Fig:4}
\end{figure}

\begin{figure}[tb]
  \SetFigLayout{2}{1}
  \centering
  \subfigure[$N_{i}$ = 4]{
  \includegraphics[width=8.0cm]{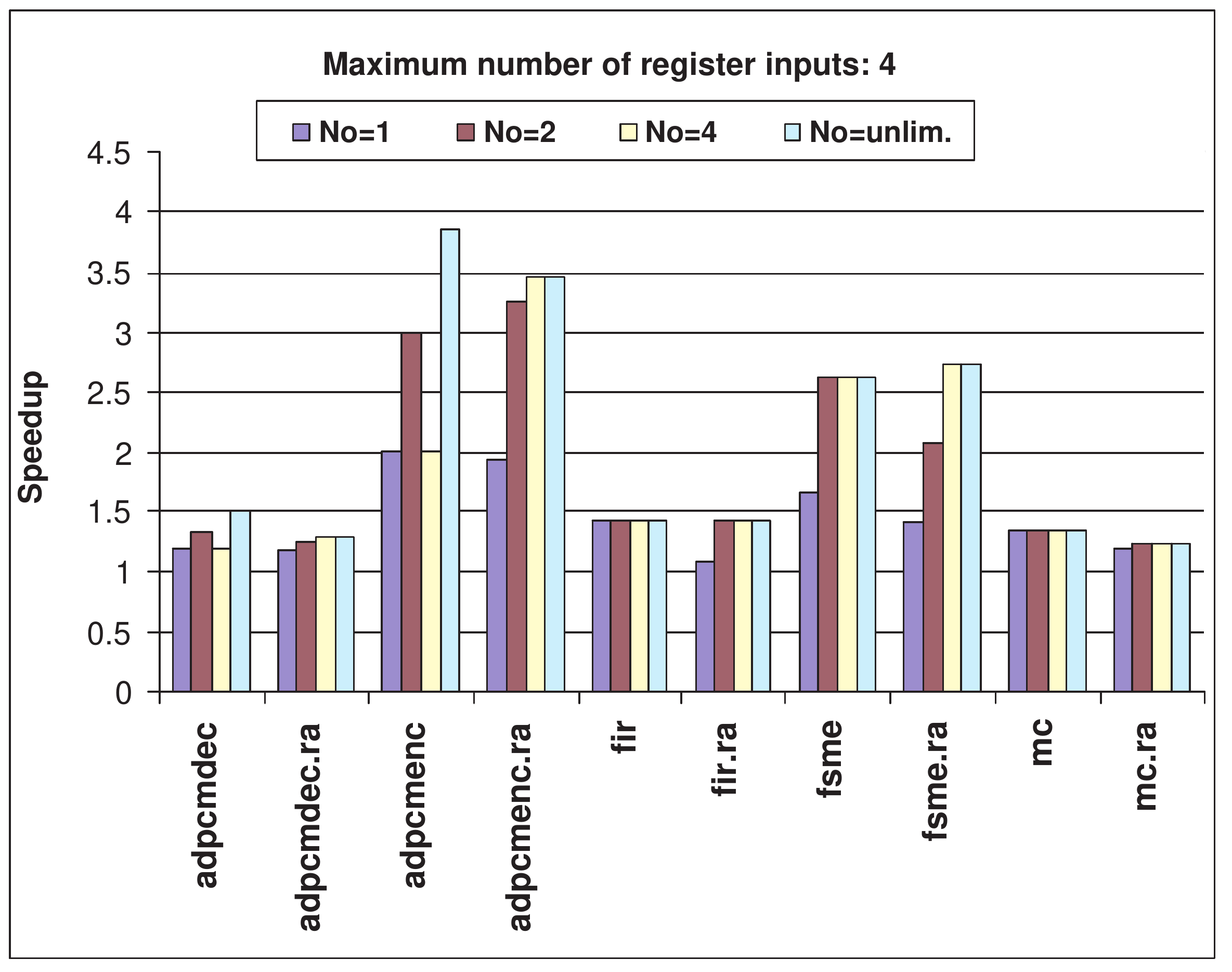}
  \label{Fig:5:a}}
  \subfigure[$N_{i}$ = 8]{
  \includegraphics[width=8.0cm]{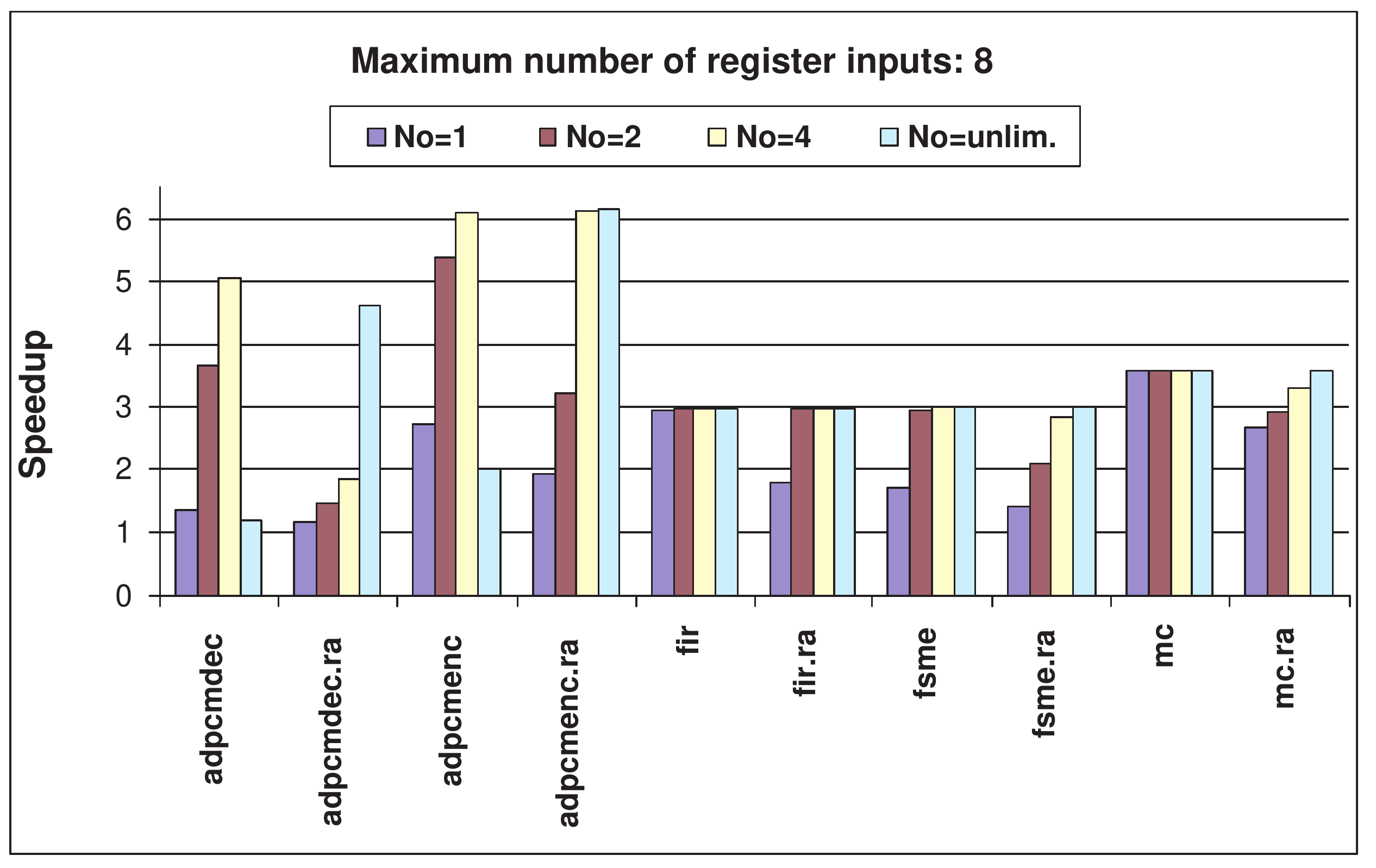}
  \label{Fig:5:b}}
  \caption{Effect of register allocation on application speedup for different number of input/output register operands.}
  \label{Fig:5}
  \vspace{-0.5cm}
\end{figure}

Application speedups obtained prior and post register allocation (the latter indicated by a `ra' suffix to the benchmark name) are shown in Fig.~\ref{Fig:5}. In contrast to common belief \cite{ClarkN03,Castro04}, the introduction of a finite register set and the mapping of the instruction selection temporaries to this set, does not always have a negative impact on the evaluated speedups. While it is clear that there is a measurable effect (an overhead of 22.15\%) due to register allocation for a single output ($N_{o}=1$), the extent of this overhead is reduced for larger number of register outputs. Thus, for $N_{o}=\{2,4,\infty\}$ the corresponding overheads have been calculated as 17\%, 2.5\% and -21.3\%, the latter meaning that the register allocated IR enables higher speedups compared to obtaining the IR prior register allocation for the constraint of unlimited number of register outputs. This important outcome infers that the overhead of spills and fills occuring due to register pressure, can be efficiently hidden when multi-output (MIMO) instructions are used for the estimations. In addition to that, CIs have the side-effect of eliminating the need for certain temporary variables within a CI pattern, given that they need not be alive outside the pattern. 

\begin{figure}[tb]
  \SetFigLayout{2}{1}
  \centering
  \subfigure[$N_{i}$ = 4]{
  \includegraphics[width=8.0cm]{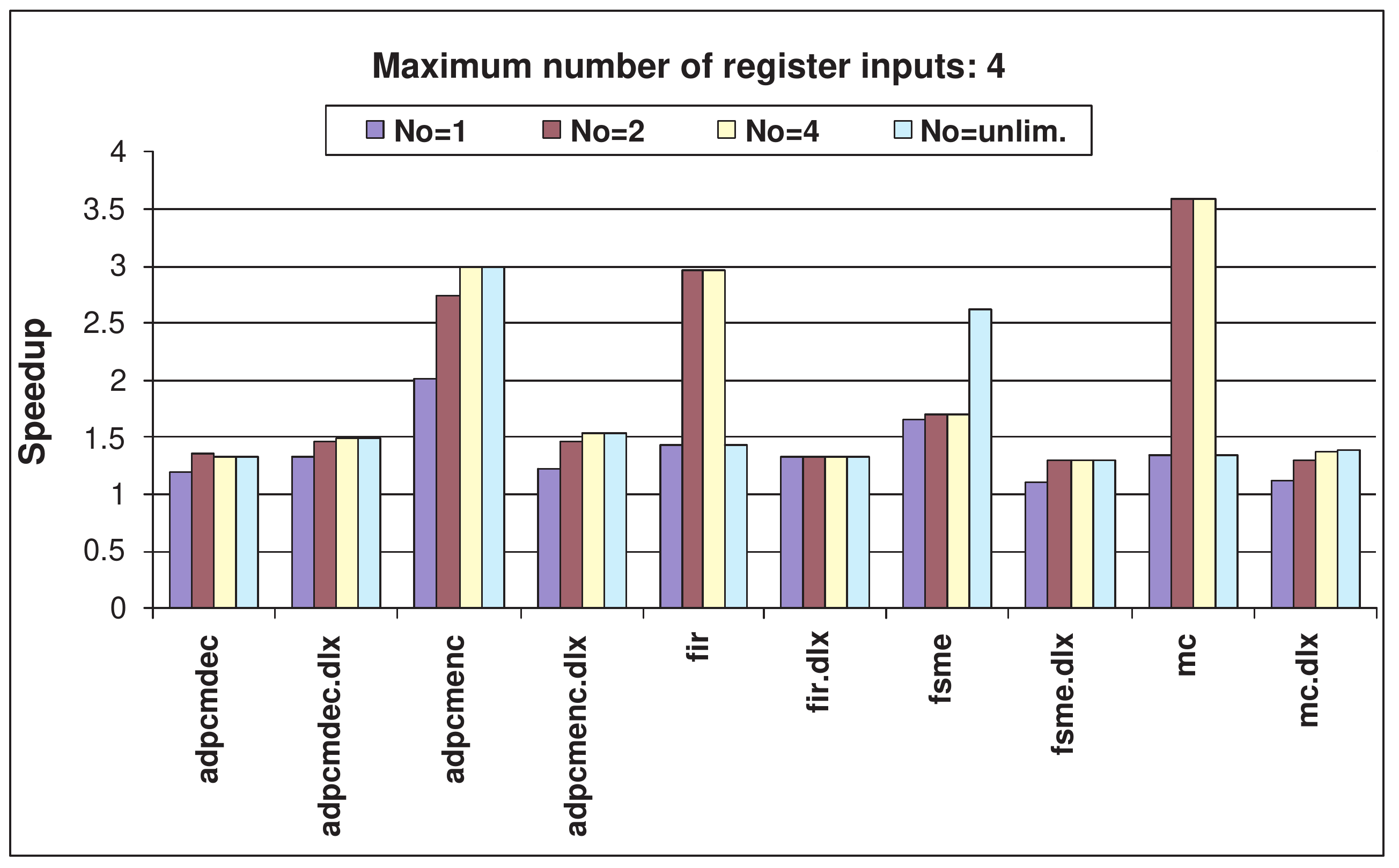}
  \label{Fig:6:a}}
  \subfigure[$N_{i}$ = 8]{
  \includegraphics[width=8.0cm]{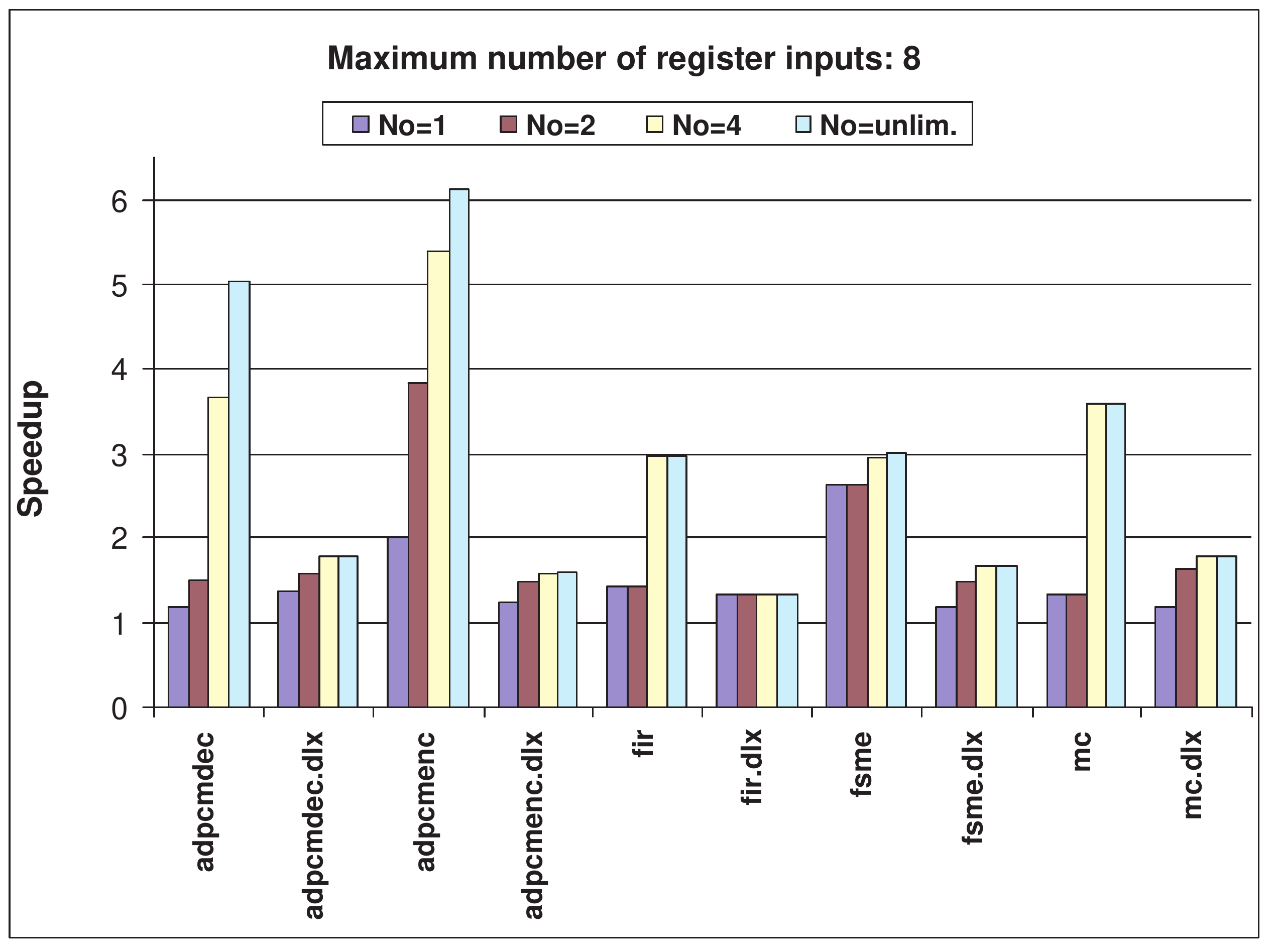}
  \label{Fig:6:b}}
  \caption{Application speedup for different number of input/output register operands on {\it SUIFvmenh} and {\it iDLX}.}
  \label{Fig:6}
  \vspace{-0.125cm}
\end{figure}

Fig.~\ref{Fig:6} shows the results on relative application speedups for different number of input/output register operands for the two selected compiler targets. An unlimited number of inputs was also set but the corresponding results where within 0.4\% of the $N_{i}=8$ case.
The difference in the average speedup achieved for the given numbers of inputs for the same application is about 44\% (ranging from 20\% to 61\%). This is partially due to the fact that stack argument allocation applied for {\it iDLX} only, adds memory access operations for saving and restoring function arguments that are not usually included in new CIs. Even when MIMO instructions are identified incorporating the callee saved sequence (a series of {\it sw} instructions), the obtained speedups are severely limited by the data memory bandwidth assumed in the estimations which is 1R/1W port for all target architectures.

\subsection{Transformation to more suitable IRs}
\label{Sec:IRTransformations}
Although not explicitly stated in related works, the effect of IR selection significantly affects the quality of the CI generation results. In YARDstick, GGX XML graph representations of ISeq patterns can be automatically generated and then transformed by hand-written AGG rules to use different IR operators for implementing equivalent functionality.

\begin{figure}[tb]
  \SetFigLayout{2}{1}
  \centering
  \subfigure[Visualization of an example host {\it SUIFvmenh} IR graph.]{
  \includegraphics[width=8.0cm]{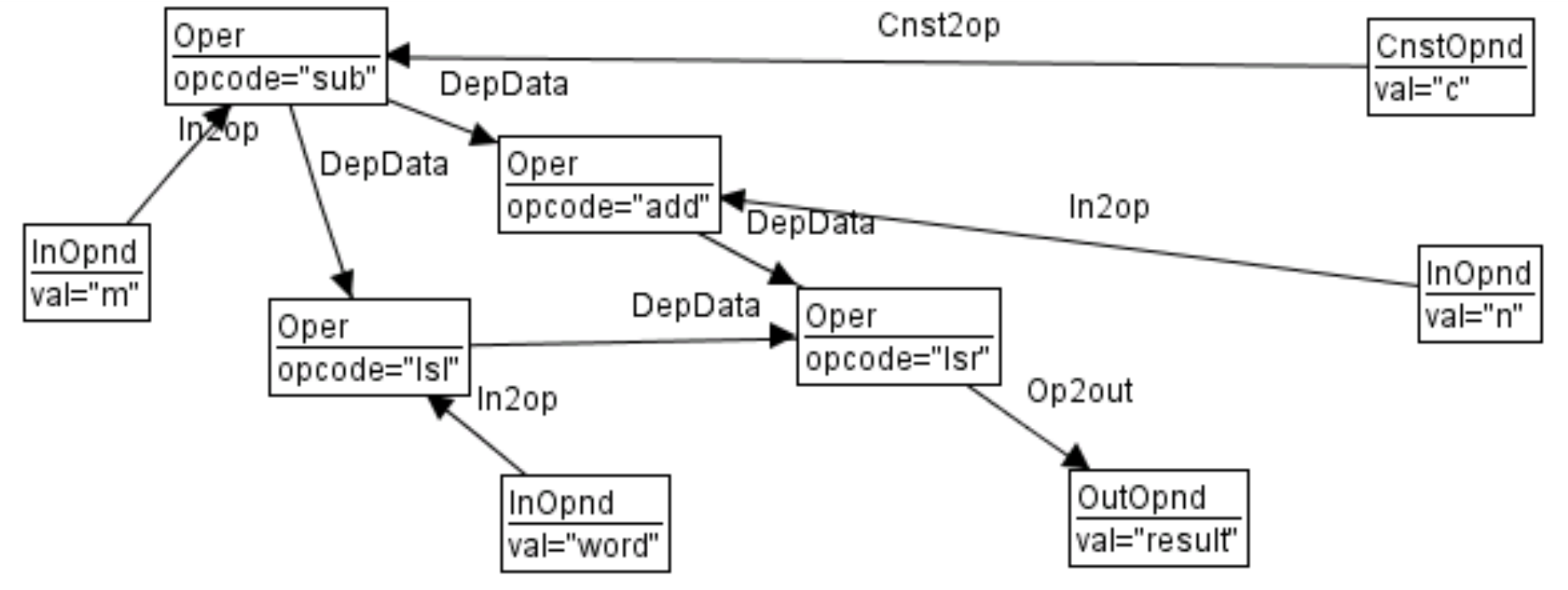}
  \label{Fig:7:a}}
  \subfigure[The resulting graph after the application of a transformation rule for `bitextract'.]{
  \includegraphics[width=8.0cm]{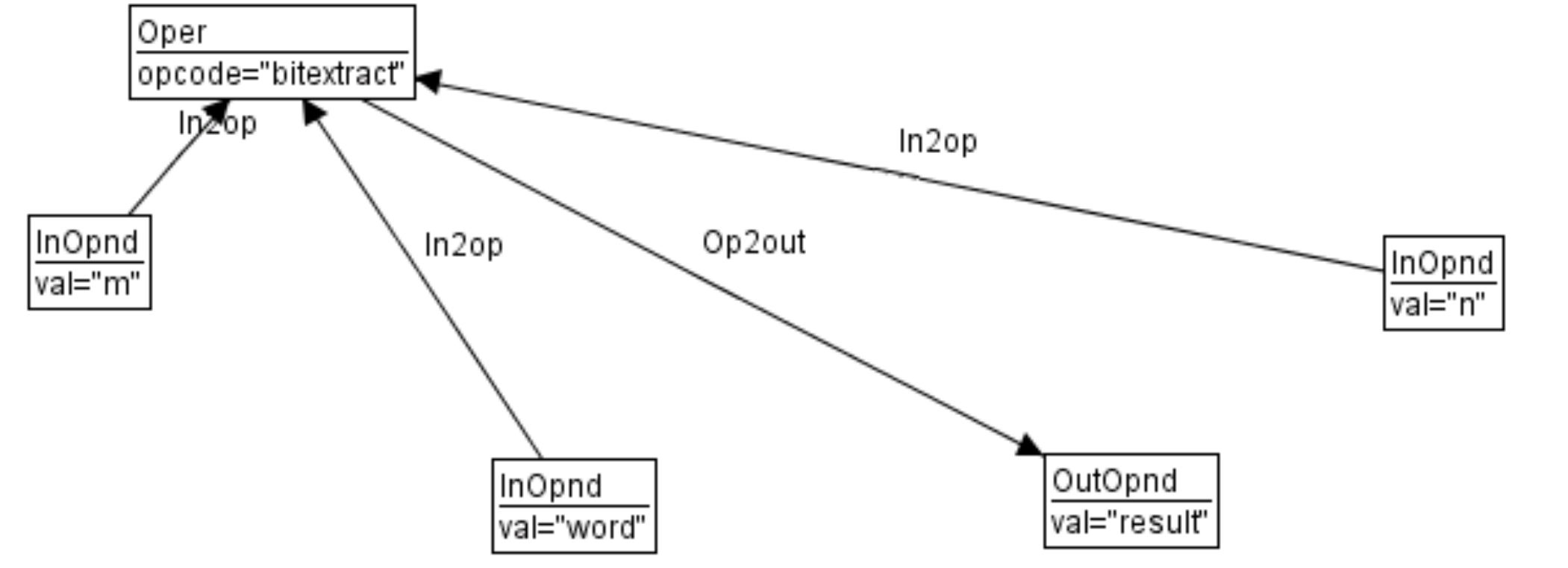}
  \label{Fig:7:b}}
  \caption{An example of IR graph rewriting via AGG transformation rules.}
  \label{Fig:7}
  \vspace{-0.125cm}
\end{figure}

\begin{figure}[tb]
  \SetFigLayout{2}{1}
  \centering
  \subfigure[The {\it crcsp}-induced CI without the use of bit-level operators.]{
  \includegraphics[width=8.0cm]{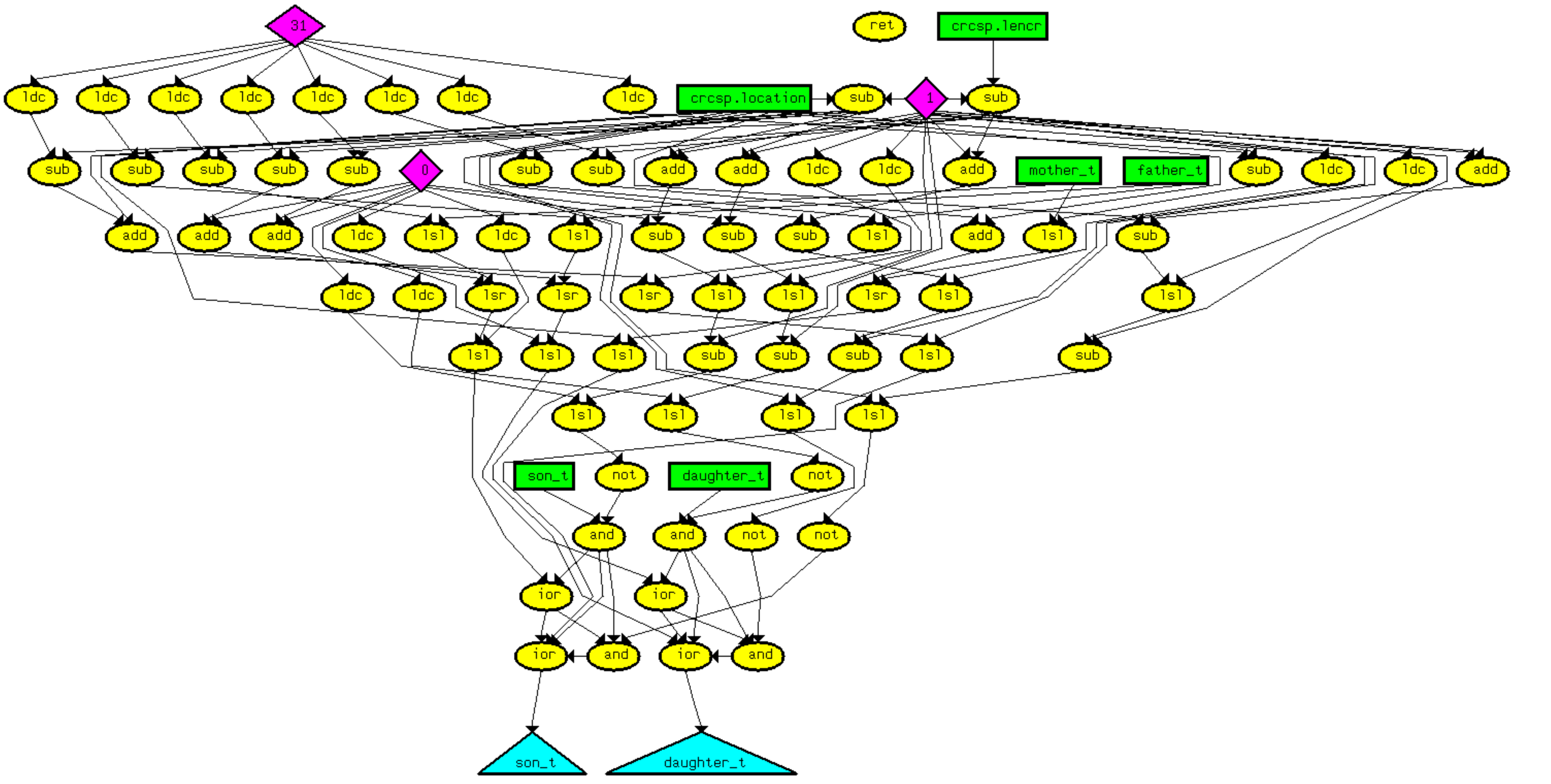}
  \label{Fig:8:a}}
  \subfigure[The {\it crcsp}-induced CI using bit-level operators.]{
  \includegraphics[width=8.0cm]{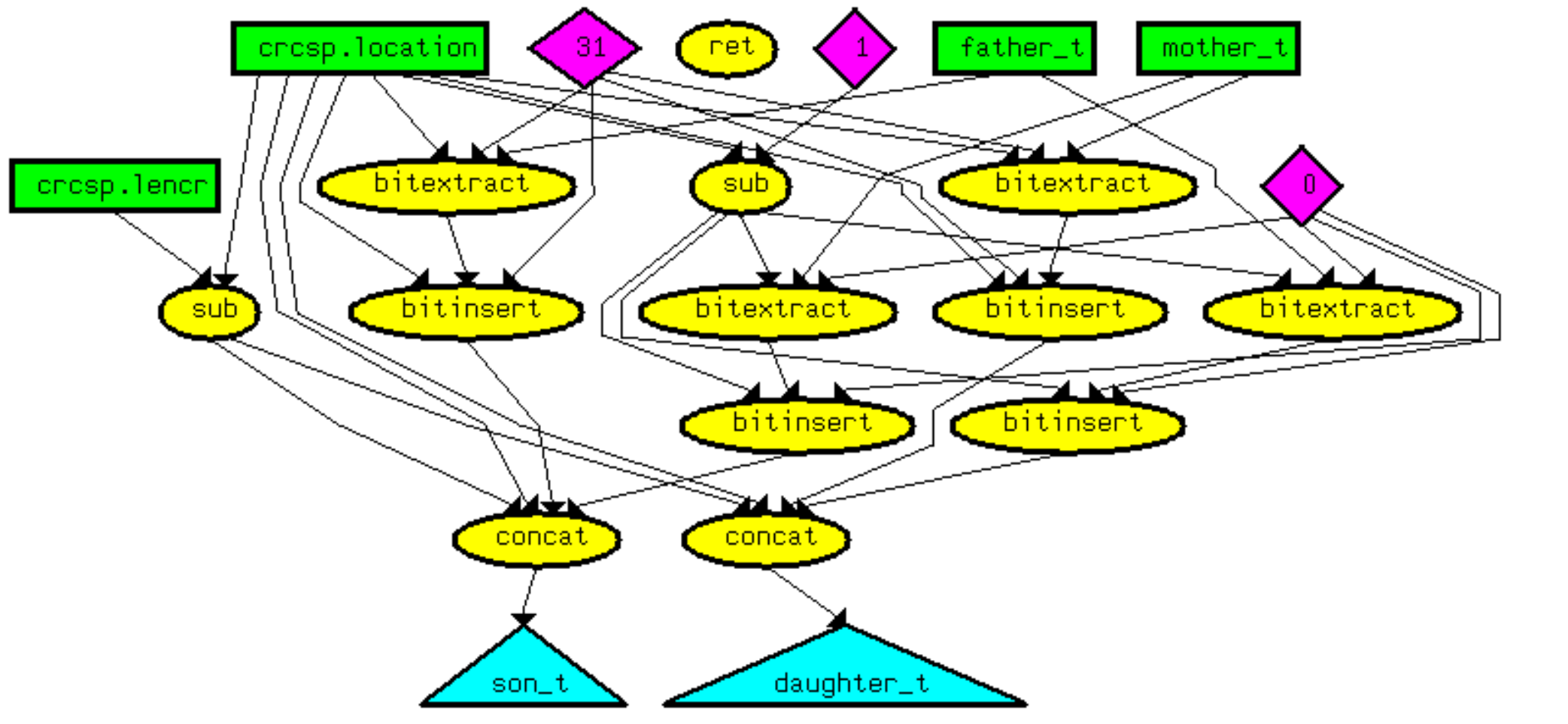}
  \label{Fig:8:b}}
  \caption{Visualization of the {\it crcsp} genetic operator CI for different compiler IRs.}
  \label{Fig:8}
  \vspace{-0.125cm}
\end{figure}

Most compilers (one exception is the commercial CoSy \cite{ACE}) do not account for bit-level manipulations that are desirable in application domains such as network processing and genetic algorithms (GAs). To highlight this issue we have defined three custom IR operators, namely {\it bitinsert, bitextract} and {\it concat} with the semantics of Table~\ref{Tab:3}. As motivational examples, we have used the well-known single- ({\it crcsp}) and double-point ({\it crcdp}) crossover operators, which are encountered in typical GAs such as the SGA \cite{Goldberg89}. It should be noted that the ANSI C implementations of the crossover operators where hand-tuned, with optimizations including the conversion of all function calls inside the {\it crcsp} and {\it crcdp} functions to macro-inclusions. Fig.~\ref{Fig:7} shows the result of applying a rule-based transformation in AGG \cite{AGG} for replacing a {\it SUIFvmenh} IR segment (Fig.~\ref{Fig:7:a}) with a use of the {\it bitextract} IR operator as seen in the resulting graph (Fig.~\ref{Fig:7:b}. To highlight the importance of the right choice of compiler IR, Fig.~\ref{Fig:8} visualizes the VCG representations of the custom instruction generated for the {\it crcsp} genetic operator, without (Fig.~\ref{Fig:8:a}) and with the use of the bit-level IR operators (Fig.~\ref{Fig:8:b}).

\begin{table}
  \renewcommand{\arraystretch}{0.95}
  \caption{Custom IR operators improving bit-level compiler support. $r_{d}$, $r_{s}$ are register operands, {\it hpos,lpos} denote a bit range, and {\it n} is the number of arguments for a variadic operator.} 
  \centering
  {\footnotesize
  \begin{tabular}{|l|c|}
    \hline
    \multicolumn{1}{|m{2.0cm}|}{\centering Operator}
    &\multicolumn{1}{m{5.0cm}|}{\centering Semantics}\\
    \hline
    {\it bitinsert} & $r_{d}[lpos..hpos] \Leftarrow r_{s}$ \\ 
    {\it bitextract} & $r_{d} \Leftarrow r_{s}[lpos..hpos]$ \\ 
    {\it concat} & $r_{d} \Leftarrow r_{s(0)} \& r_{s(1)} \& \ldots \& r_{s(n-1)}$ \\
    \hline
  \end{tabular}
  }
  \label{Tab:3}
  \vspace{-0.125cm}
\end{table}

The performance gains for the generated hardware depend heavily on the target IR used for mapping the application code as can be clearly seen by the results of Table~\ref{Tab:4}. In Table~\ref{Tab:4}, the first three columns are self-explanatory. Column `Cycles...' shows the cycles required for a sequential schedule of the corresponding GA operator assuming the usage of the generated CIs. The last two columns indicate the number of cycles and area of the CI. The area requirement is calculated relatively to the area (multiplier area unit or MAU) of a 32-bit single-cycle multiplier producing a 64-bit result.

For computing schedules with unlimited resources, the generated ISeq files of the custom instructions were automatically converted with YARDstick to CDFGs compatible with an extended version of the CDFG toolset \cite{CDFGtool} and processed by an ASAP scheduler. If the bit-level operators are not used, the minimum number of cycles required for the {\it crcsp} operator are 76 for a sequential schedule and 12 for scheduling with unlimited resources, while for the {\it crcdp} these limits are 111 and 14, respectively. When the bit-level operators are used, the sequential schedules prior to the inclusion of custom instructions require 13 and 18 cycles for {\it crcsp} and {\it crcdp} respectively with an ASAP schedule of 5 cycles for both. In the latter case, a single-cycle MIMO custom instruction is identified for each genetic operator when a $N_{i}/N_{o}=\{8/2\}$ constraint is used with impressive area benefits as well. 

\begin{table}
  \renewcommand{\arraystretch}{0.975}
  \caption{CI characteristics for hand-optimized ANSI C implementations of {\it crcsp} and {\it crcdp}.} 
  \centering
  {\footnotesize
  \begin{tabular}{|l|l|r|r|r|r|}
    \hline
    \multicolumn{1}{|m{1.0cm}|}{\centering \footnotesize GA operator}
    &\multicolumn{1}{m{1.0cm}|}{\centering \footnotesize Bit-level operations}
    &\multicolumn{1}{m{1.25cm}|}{\centering \footnotesize CI gen. constraints \\ $N_{i}/N_{o}$}
    &\multicolumn{1}{m{1.25cm}|}{\centering \footnotesize Cycles \\ (seq. schedule)}
    &\multicolumn{1}{m{0.75cm}|}{\centering \footnotesize CI cycles}
    &\multicolumn{1}{m{1.0cm}|}{\centering \footnotesize CI area (MAU)}\\
    \hline
    {\it crcsp} & No & 4/1 & 76 & -- & -- \\ 
    {\it crcsp} & No & 8/1 & 41 & 3 & 0.977 \\ 
    {\it crcsp} & No & 8/2 & 5 & 3 & 1.867 \\ 
    \hline
    {\it crcsp} & Yes & 4/1 & 13 & -- & -- \\ 
    {\it crcsp} & Yes & 8/1 & 6 & 1 & 0.142 \\ 
    {\it crcsp} & Yes & 8/2 & 1 & 1 & 0.153 \\ 
    \hline
    {\it crcdp} & No & 4/1 & 111 & -- & -- \\ 
    {\it crcdp} & No & 8/1 & 58 & 3 & 1.466 \\ 
    {\it crcdp} & No & 8/2 & 5 & 3 & 2.800 \\ 
    \hline
    {\it crcdp} & Yes & 4/1 & 18 & -- & -- \\ 
    {\it crcdp} & Yes & 8/1 & 8 & 1 & 0.147 \\ 
    {\it crcdp} & Yes & 8/2 & 1 & 1 & 0.164 \\ 
    \hline
  \end{tabular}
  }
  \label{Tab:4}
  \vspace{-0.25cm}
\end{table}

\subsection{Effect of data memory access model}
\label{Sec:MemoryAccessModel}
The extent and scope of using custom instructions is constrained by the data bandwidth to the data memory unit and local storage (register file) as defined by the number of input/output ports and the resolution of dependencies for load/store operations. In certain approaches \cite{ClarkN05,Leupers06} which deal with predefined architectures such as the MIPS CorExtend and ARM OptimoDE systems, this limitation imposes a definitive factor. However, for exploration purposes when developing an ASIP from scratch, it is useful to consider different storage consistency models. Following the notation introduced in \cite{Biswas07} for state consistency between application-specific functional units (AFUs) with local storage and data memory, it is possible to consider two such models in YARDstick:
\begin{itemize}
\item {{\it Consistent data memory}, where the AFU directly accesses data in the on-chip data memory and there is no need for local AFU storage. We also make the conservative assumption that loads and stores ought always be serialized.}
\item {{\it Ideal consistent AFU memory}, where each load/store to main memory is transformed to an access to local AFU memory. We assume that data memory status is updated by DMA accesses occuring in parallel to processor instructions.}
\end{itemize}   

\begin{figure}[tb]
  \SetFigLayout{2}{1}
  \centering
  \subfigure[$N_{i}/N_{o}$ = 4/2]{
  \includegraphics[width=8.0cm]{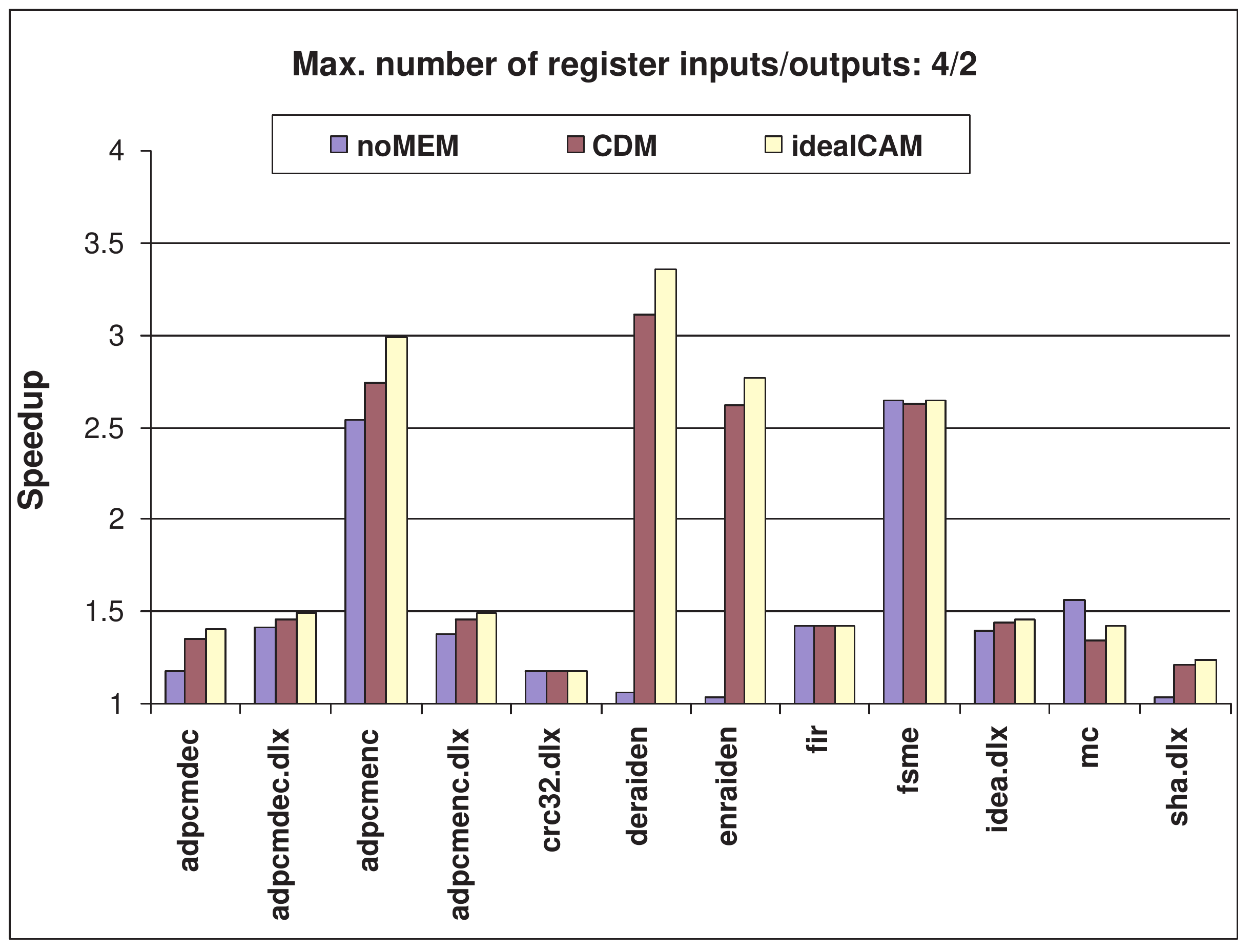}
  \label{Fig:9:a}}
  \subfigure[$N_{i}/N_{o}$ = 8/4]{
  \includegraphics[width=8.0cm]{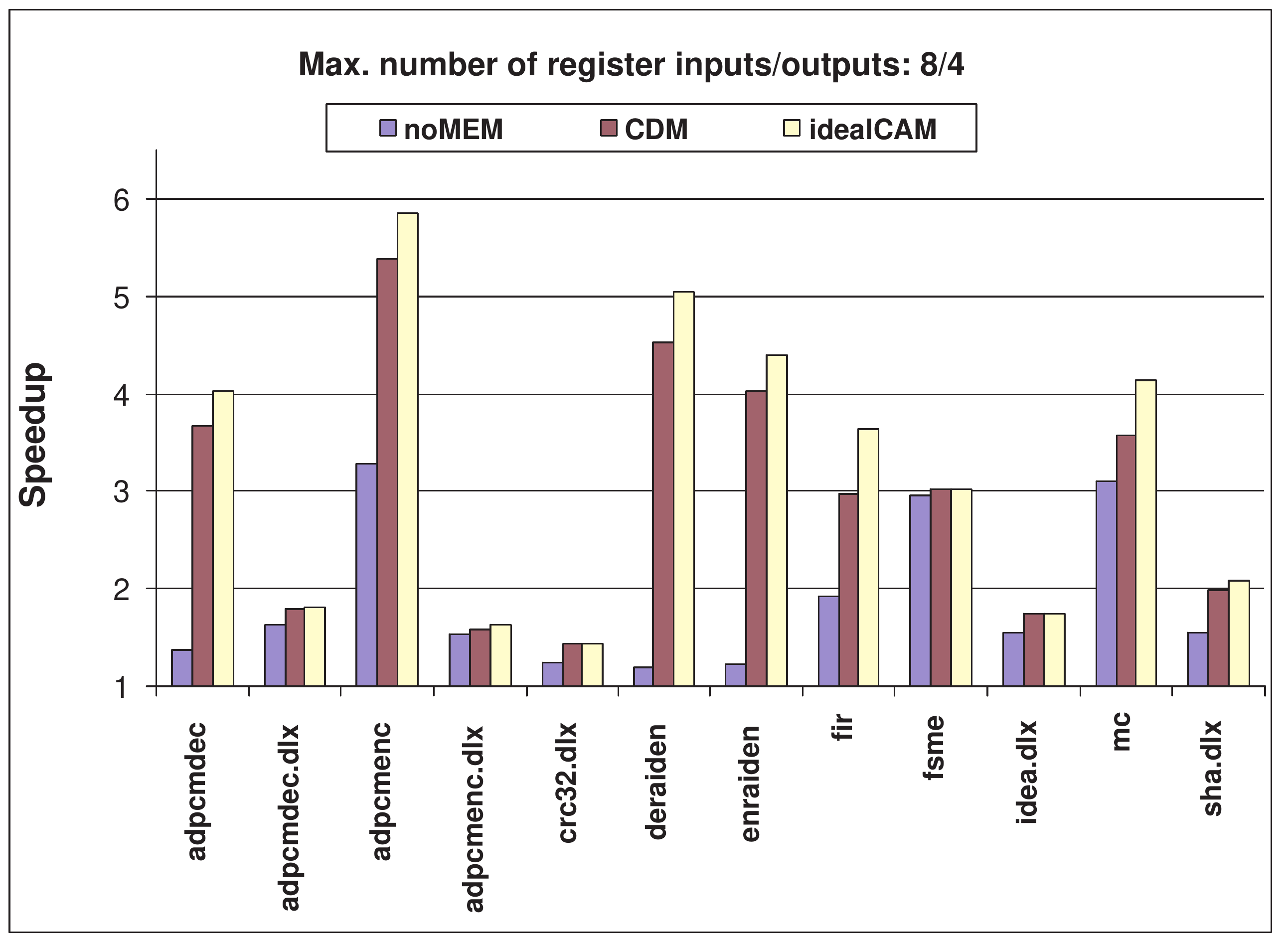}
  \label{Fig:9:b}}
  \caption{Effect of data memory accesses to the speedup induced by CIs. Accesses to data memory are assumed to require a single clock cycle overhead.}
  \label{Fig:9}
  \vspace{-0.25cm}
\end{figure}

To investigate the effect of memory model choice on application speedup due to CIs we first generated CIs without allowing memory inclusion (``noMEM''), then allowed local memory and performed estimations for the consistent data memory model (``CDM'') and subsequently we assumed an ideal consistent AFU memory (``idealCAM''). The corresponding results are illustrated in Fig.~\ref{Fig:9} for indicative $N_{i}/N_{o}$ combinations and for a single-issue processor. 

As can be seen by the collected results, the inclusion of data memory access operations in CIs has a significant positive impact in the attained speedups: from 15.5\% to 33.3\% for the given input/output constraints. Especially for the {\it SUIFvmenh} target, the speedup improvements were up to 43.4\%. Another important observation is that the consistent AFU memory model has a limited effect with improvements of up to 6.3\% in average and 8.9\% for the {\it SUIFvmenh} applications alone. However, for a larger cycle overhead to accessing data storage, the speedup improvements are more considerable. For another exploration example, we have estimated that 2- and 5-cycle load/store accesses to a data memory module through the local bus (an address cycle followed by either one word data access or consecutive byte data access cycles) result in higher speedups. More specifally, the ``CDM'' case performs better to ``noMEM'' by 34.7\% and 49.9\%, respectively for the 2- and 5-cycle overheads. When comparing the two different models that allow memory accesses to be part of CIs, the corresponding values are 7\% and 20.9\% in favor of ``idealCAM'' for the given cycle overheads. 

\subsection{Greedy CI selection under priority metrics}
\label{Sec:GreedySelection}
For implementing a greedy CI selector, the key idea is to assign priorities to the CI patterns and the more proficient instances are chosen by starting with the highest prioritized one. We have used the following two priority functions: 

\begin{equation}
\label{Eq:1}
\text{Cycle gain}: Priority(\sum_{j} C_{i,j}) = \sum_{j} \{P_{i,j} \times f_{i,j}\}
\end{equation}

that forces for best performance regardless AFU area requirements and: 

\begin{equation}
\label{Eq:2}
\text{Cycle gain/Area}: Priority(\sum_{j} C_{i,j}) = \sum_{j} \{(P_{i,j} \times f_{i,j})\}/A_{i}
\end{equation}

where $C_{i,j}$ denotes the $i$-th candidate instruction with $j$ different instances in the entire program, $f_{i,j}$ the basic block execution frequency metric associated with the specific instance, and $A_{i}$ the area cost for the candidate. These priority functions force different objectives: equation~\ref{Eq:1} maximizes performance gain for each isomorphic candidate CI over the entire program when area is not an issue while equation~\ref{Eq:2} quantifies the available area budget as well.

A summary of the measurements for the application set is given in Table~\ref{Tab:5}. 
Taking {\it sha.dlx} for example, although tens of candidate instructions are identified, 
only a few (7 for achieving 95\% of the maximum speedup compared to 20 for achieving totality) contribute significantly to the execution time for either priority function. The 
number of required extensions for reaching the 95\% speedup levels ranges from 2 ({\it fir.dlx}) to 40 ({\it idea.dlx}), while the area requirement is less than 3.4 multiples of the area of a 32-bit single-cycle multiplier for all applications with the exception of {\it idea.dlx} which demands up to 10.23 MAU. 

\begin{figure}[tb]
  \centering
  \includegraphics[width=8.0cm]{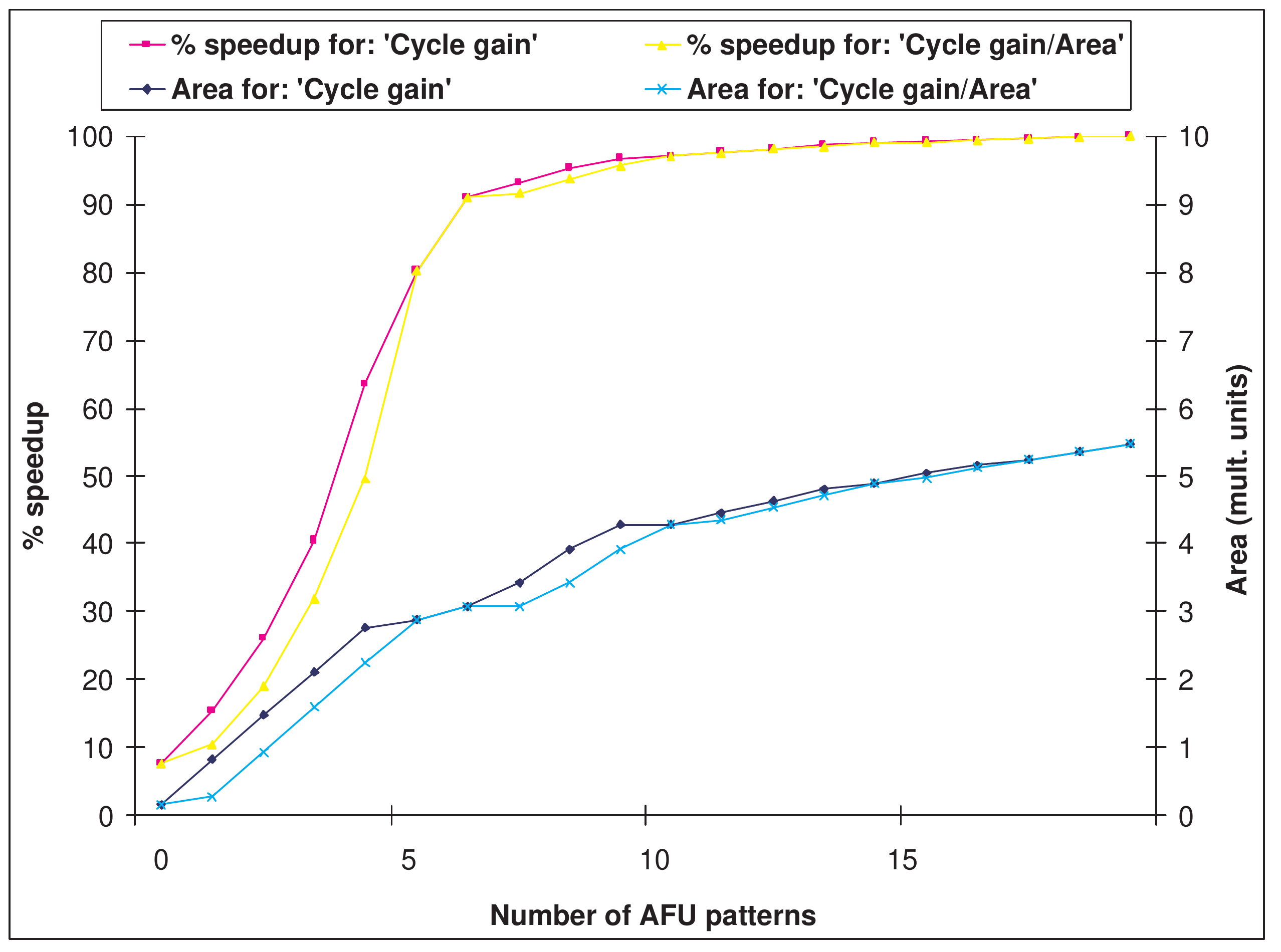}
  \caption{Custom instruction selection under priority metrics for {\it sha}
  ($N_{i}/N_{o}=\infty/\infty$).}
  \label{Fig:10}
  \vspace{-0.175cm}
\end{figure}

Finally, Fig.~\ref{Fig:10} compares the pros and cons for the priority functions 
used in the custom instruction selection process for the {\it sha.dlx} application example. For the {\it sha} application, CI selection under the `Cycle gain' priority function reaches the 95\% of the maximum speedup for one instruction less and a slight area increase (0.04 MAU) compared to `Cycle gain/Area'.

\begin{table}
  \renewcommand{\arraystretch}{1.0}
  \caption{Speedup-AFU area for `Cycle gain'/`Cycle gain/Area' 
  for the input/output constraint $N_{i}/N_{o}=\{8/4\}$.} 
  \centering
  {\footnotesize
  \begin{tabular}{|l|r|r|r|r|}
    \hline
    \multicolumn{1}{|m{1.2cm}|}{\centering Benchmark}
    &\multicolumn{1}{m{1.0cm}|}{\centering 0.95$\times$ max. speedup}
    &\multicolumn{1}{m{1.2cm}|}{\centering Area (MAU)}
    &\multicolumn{1}{m{1.0cm}|}{\centering At max. speedup}
    &\multicolumn{1}{m{1.2cm}|}{\centering Area (MAU)}\\
    \hline
    {\it adpcmdec} & 4/4 & 0.895/0.895 & 6 & 0.983 \\ 
    \hline
    {\it adpcmdec.dlx} & 11/11 & 1.123/1.123 & 17 & 1.721 \\ 
    \hline
    {\it adpcmenc} & 4/4 & 0.998/0.998 & 6 & 1.086 \\
    \hline
    {\it adpcmenc.dlx} & 16/16 & 1.475/1.375 & 22 & 2.074 \\
    \hline
    {\it crc32.dlx} & 3/3 & 0.12/0.12 & 3 & 0.12 \\
    \hline
    {\it deraiden} & 4/4 & 2.657/2.657 & 4 & 2.657 \\
    \hline
    {\it enraiden} & 3/3 & 1.949/1.949 & 3 & 1.949 \\
    \hline
    {\it fir} & 4/4 & 1.398/1.398 & 5 & 1.398 \\
    \hline
    {\it fir.dlx} & 2/2 & 0.155/0.155 & 2 & 0.155 \\
    \hline
    {\it fsme.dlx} & 9/9 & 1.143/1.143 & 11 & 1.546 \\
    \hline
    {\it fsme.dlx} & 6/6 & 1.03/1.03 & 10 & 1.65 \\
    \hline
    {\it idea.dlx} & 40/50 & 10.23/9.325 & 69 & 13.002 \\
    \hline
    {\it mc} & 5/5 & 1.824/1.824 & 7 & 2.53 \\
    \hline
    {\it mc.dlx} & 7/7 & 1.489/1.489 & 12 & 2.516 \\
    \hline
    {\it sha.dlx} & 7/7 & 1.671/1.671 & 20 & 3.378 \\
    \hline
  \end{tabular}
  }
  \label{Tab:5}
  \vspace{-0.25cm}
\end{table}

\section{Usage environment}
\label{Sec:Usage}
YARDstick has been used along with the SUIF/Machine-SUIF \cite{MachSUIF}, GCC \cite{GCC}, and COINS \cite{COINS} compilers and the ArchC \cite{ArchC} simulation framework. Functional and cycle-accurate simulators generated by version 1.5.1 of ArchC can be used with YARDstick without any modifications. Most of the YARDstick functionality is also accessible through a cross-platform GUI \cite{Kavvadias07} compatible to recent Tcl/Tk versions (8.5.a5 and newer). 

Supported platforms include GNU/Linux (RedHat 9.0), Cygwin and Win32 (Windows/XP SP2) on x86-compatible processors.

\section{Conclusions}
\label{Sec:Conclusions}
YARDstick is a retargetable application analysis and custom instruction generation/selection environment providing a compiler-/simulator-agnostic infrastructure. YARDstick aims in separating design space exploration from compiler/simulator idiosyncrasies. Different compilers/simulators can be plugged-in via file-based interfaces; further, both high- (e.g. ANSI C) and low-level (assembly for an architecture or a virtual machine) input can be analyzed by the infrastructure.

In order to prove the applicability and usefulness of YARDstick in ASIP development, we have evaluated a variety of exploration scenarios on a benchmark set consisting of well-known embedded applications and kernels. In this context, we have investigated effects of the compilation process, such as the selection of the target IR and the impact of register allocation, on the characteristics of the identified hardware extensions. Also, different memory models involving local storage for application-specific functional units were examined and quantified, and for the entire set of applications, custom instructions were generated under different input/output constraints.



\nocite{*}

\bibliographystyle{compj}
\bibliography{yardstick}

\end{document}